\documentclass[aps,prstab,preprint,groupedaddress]{revtex4}
\usepackage{graphics,epsfig}
\begin{document}

\draft
\title{Theory of Networked Minority Games based on Strategy Pattern
Dynamics}
\author{T. S. Lo,$^{1}$ H. Y. Chan,$^{1}$ P. M. Hui,$^{1}$ and N. F.
Johnson$^{2}$}
\address{$^{1}$Department of Physics, The Chinese University of
Hong Kong\\
Shatin, New Territories, Hong Kong \\
$^{2}$Department of Physics, University of Oxford\\
Clarendon Laboratory, OX1 3PU, United Kingdom}

\begin{abstract}
We formulate a theory of agent-based models in which agents
compete to be in a winning group. The agents may be part of a
network or not, and the winning group may be a minority group or
not.  The novel feature of the present formalism is its focus on
the dynamical pattern of strategy rankings, and its careful
treatment of the strategy ties which arise during the system's
temporal evolution. We apply it to the Minority Game (MG) with
connected populations.  Expressions for the mean success rate
among the agents and for the mean success rate for agents with $k$
neighbors are derived.  We also use the theory to estimate the
value of connectivity $p$ above which the Binary-Agent-Resource
system with high resource level goes into the high-connectivity
state.

\noindent PACS Nos.: 02.50.Le, 05.65.+b, 05.40.-a, 89.90.+n

\end{abstract}
\maketitle

\section{Introduction}

Agent-based models form an important part of research on complex
adaptive systems \cite{recentactivities}.  For example, the
self-organization of an evolving population consisting of agents
competing for a limited resource has potential applications in
areas such as economics, biology, engineering, and social sciences
\cite{recentactivities,ourbook}.  The bar-attendance problem
proposed by Arthur \cite{arthur1,johnson1} constitutes a typical
setting of such a system in which a population of agents decide
whether to go to a popular bar having limited seating capacity.
The agents are informed of the attendance in past weeks, and hence
the agents share common information, interact through their
actions, and learn from past experience. The problem can be
simplified by considering binary games, either in the form of the
minority game (MG) \cite{challet1,challet0} or in a
binary-agent-resource (B-A-R) game \cite{johnson2,zheng2}. For
modest resource levels in which there are more losers than
winners, the minority Game proposed by Challet and Zhang
\cite{challet1,challet7} represents a simple, yet non-trival,
model that captures many of the essential features of such a
competing population.

The MG considers an odd number $N$ of agents. At each time step,
the agents independently decide between two options `$0$' and
`$1$'. The winners are those who choose the minority option.  The
agents learn from past experience by evaluating the performance of
their strategies, where each strategy maps the available global
information (i.e., the record of the most recent $m$ winning
options) to an action. One important quantity in MG is the
standard deviation $\sigma$ of the number of agents making a
particular choice. This quantity reflects the performance of the
population as a whole in that a small $\sigma$ implies on average
more winners per turn, and hence a higher success rate per turn
per agent. In the MG, $\sigma$ exhibits a non-monotonic dependence
on the memory size $m$ of the agents
\cite{savit,challet3,challet4}. When $m$ is small, there is
significant overlap between the agents' strategies. This crowd
effect \cite{johnson2,johnson3,hart1} leads to a large $\sigma$,
implying the number of losers is high. This is the crowded, or
informationally efficient, phase of MG. In the informationally
inefficient phase where $m$ is large, $\sigma$ is moderately small
and the agents perform better than if they were to decide their
actions randomly.

Theoretical analysis of the MG has been the focus of many studies
\cite{johnson1,challet3,challet4,johnson2,johnson3,hart1,challet5,challet6,coolen1,coolen2,challet8,galla}.
Mapping the MG into the language of disordered spin systems makes
the machinery in statistical physics of disordered system, most
noticeably the replica trick, useful in the study of models of
competing population.  For the MG, calculations based on the
replica method work well for the case of large strategy pool,
i.e., when the strategy pool is much larger than the strategies
actually being used in making decisions.  This is referred to as
the informationally inefficient phase because information is left
in the resulting bit-string patterns for a single realization of
the game. In the informationally efficient phase, the whole pool
of strategies tend to be in play during the game. The
crowd-anticrowd theory gives a physically transparent quantitative
theory of the observed features in this regime, as well as in the
inefficient regime. The crowd-anticrowd theory is based on the
fact that it is the difference in the numbers of agents playing a
strategy $R$ and the corresponding anti-correlated strategy $R'$
that plays the most important role in the understanding of the
fluctuations and hence performance of the whole population.

The crowd-anticrowd theory is a microscopic approach in the sense
that it follows the strategy play of the agents in the population.
While this leads to microscopically correct equations, in practice
these equations are evaluated by simply time-averaging over the
path taken by the strategy rankings. A naive time-averaging over
all histories becomes more difficult to implement as the number of
ties in strategy scores increases, since such ties affect the
number of agents playing a given ranking of strategy and hence
must be incorporated explicitly in the time-average. The effect of
strategy ties becomes increasingly important as $m$ decreases,
i.e., size of strategy space decreases, since the chance of ties
arising will then increase. In the efficient phase at low $m$,
therefore, there are frequent ties in the strategy performance
\cite{johnson1,gourley}, and hence the time-averagings within the
crowd-anticrowd theory require special care. In the present paper,
we present a complementary theoretical treatment for this regime
in order to explicitly account for such strategy ties. The
resulting theory amounts to a non-trivial reorganization of the
time-averagings within the crowd-anticrowd theory. Like the
original crowd-anticrowd theory, it is applicable to both
non-networked and networked populations \cite{gourley,choe}. The
theory we present is based on the idea of following the patterns
of the ranking of strategies as the game evolves in time, without
knowing the details on the ranking of each strategy. The effects
of tied strategies are taken into account by considering the
number of strategies belonging to each rank as the game evolves.
To illustrate the applications of the theory, we show that the
theory explains the features observed in numerical results of a
networked version of MG.  The theory also allows the evaluation of
the success rate of agents with a given number of connected
neighbors. The latter is important in the study of the
functionality of an underlying network \cite{mendes} in a
competing population.

The plan of the paper is as follows.  In Sec.II, we define the MG
in non-networked and networked populations.  In Sec.III, we
discuss the different ranking patterns based on performance of the
strategies as the game evolves, and the fraction of strategies in
each rank.  The number of agents using a strategy belonging to a
particular rank, is derived for both non-networked and networked
MG in Section IV.  In Sec.V,  we apply the theory to derive an
expression for the mean success rate in the efficient phase as a
function of the connectivity in the population and compare results
with those obtained by numerical simulations.  An alternative way
to study the mean success rate is to decompose the population into
agents with different numbers of connected neighbors.  An
expression for the mean success rate of agents with $k$ connected
neighbors is derived in Sec.VI.  Results are in agreement with
numerical simulations.  Section VII gives a discussion on the
limit of validity of the theory and shows that the theory can be
extended to study B-A-R models at high connectivity.

\section{The Minority Game}

The basic MG \cite{challet1} comprises of $N$ agents competing to
be in a minority group at each time step. The only information
available to the agents is the history. The history is a
bit-string of length $m$ recording the minority option for the
most recent $m$ timesteps. There are a total of $2^{m}$ possible
history bit-strings. For example, $m=2$ has $2^2=4$ possible
global outcome histories: $00$, $01$, $10$ and $11$. At the
beginning of the game, each agent picks $s$ strategies, with
repetition allowed. They make their decisions based on their
strategies. A strategy is a look up table with $2^{m}$ entries
giving the predictions for all possible history bit-strings. Since
each entry can either be `0' or `1', the full strategy pool
contains $2^{2^{m}}$ strategies. Adaptation is built in by
allowing the agents to accumulate a merit (virtual) point for each
of her strategies as the game proceeds, with the initial merit
points set to zero for all strategies. Strategies that predicted
the winning (losing) action at a given timestep, are assigned
(deducted) one (virtual) point. At each turn, the agent follows
the prediction of her best-scoring strategy.  A random choice will
be made for tied strategies.

The networked Minority Game \cite{gourley,choe} explores the {\em
functionality} of an underlying network in the context of a
population competing for limited resource.  At each timestep $t$,
each agent (node) decides on one or two options, as in the basic
MG. Each agent decides in light of (i) {\em global information}
which takes the form of the history of the $m$ most recent global
outcomes as in the basic MG, and (ii) {\em local information}
obtained via network connections. The connections here need not be
physical -- it  only matters that the connected neighbors are
those with whom an agent can communicate. Adaptation is introduced
by randomly assigning $s$ strategies to each agent.  At each
timestep, each agent compares the score of its own best-scoring
strategy (or strategies) with the highest-scoring strategy (or
strategies) among the agents to whom he is connected. The agent
adopts the action of whichever strategy is highest-scoring
overall, using a coin-toss to break any ties. The network can be a
classical random network or take on the geometry of a growing
scale-free network \cite{mendes}. For simplicity, we here assume a
random network, where the connection between any two agents (i.e.,
nodes) exists with a probability $p$. Numerical results
\cite{gourley} show that the presence of connections lowers the
global performance of the population, while ensuring fairness by
lowering the spread in the success rates among the agents.  Here,
we aim at formulating a theory that can be applied to explain the
features observed in the numerical simulations. Since the effect
of such connections is typically to increase the chances of
strategy ties, particularly at low $m$, this motivates the present
theory's approach of tracking strategy patterns in time.

\section{Ranking the strategies}

The two key ingredients for the present theory are (i) the patterns
of strategy rankings according to performance, based on the strategies'
virtual points as the game
proceeds; and (ii) the fraction of strategies in each rank for
each ranking pattern. In the following two subsections, we discuss
these two points. The discussion is valid for populations either with
or without connections.

\subsection{Ranking Pattern}

As a particular run of a given game evolves, the pattern of
strategy rankings also evolves.  The instantaneous strategy
ranking depends on the number of history bit-strings that have
occurred an {\em odd} number of times and the next outcome will
depend on whether the current history bit-string has occurred an
odd or even number of times.  Both of these factors are important
in the calculation of the mean success rate of the population.

Suppose we are at a given moment in the run of a game.  Let $\mu$
be the current history bit-string that the agents are using for
decisions. Let $\{t_{odd}^{\nu}\}$ be the set of turns (i.e.,
timesteps) so far in which a history $\nu$ has occurred an odd
number of times (including the initial history bit-string that
starts the game) while $\{t_{even}^{\nu}\}$ is the set of turns so
far in which a history $\nu$ has occurred an even number of times
\cite{larry1}. For small values of $m$, i.e. in the efficient
phase of the MG, the outcome time series exhibits the feature of
anti-persistence or double periodicity
\cite{savit,challet3,challet4,zheng1,Jefferies1}. This feature
implies that all history bit-strings occur with equal
probabilities. It means that for a current history $\mu$ based on
which that the agents decide, if the winning side is $\eta$
($\eta$ can be $0$ or $1$) when $t\in \{t_{even}^{\mu}\}$, the
outcome is $1-\eta $ with probability unity in the next occurrence
of $\mu$. This follows that no strategies could perform better
than the others in an average over time, and the virtual points
(VP) of the strategies cannot show a runaway behavior, i.e., the
VPs of strategies will not keep on increasing or decreasing. This
property is intimately related to the fact that the Eulerian Trail
is an underlying quasi-attractor of the game in this efficient
regime \cite{Jefferies1}. By focusing on whether a history has
occurred an odd or even number of times during a run, we are
picking out what is essentially the most important aspect of the
outcome series.

For a particular turn $t$, we define the ranking of the strategies
according to their performance up to that point in time based on
the VPs of the strategies.  The Rank-$1$ strategy or strategies
have the highest VPs. The Rank-$2$ strategies are the second
best-performing (having the second highest VPs), and so on. For
small $m$ (efficient phase), the ranking pattern of the strategies
depends on {\em the number of histories} that have occurred an odd
number of times.  It is illustrative to consider an example for
the case of $m=2$ where there are $4$ possible histories ($00$),
($01$), ($10$), ($11$). At $t=0$, all strategies are assigned the
same VP. There is only one rank, called Rank-$1$, of the
strategies, with all the strategies belonging to this rank. This
is also the case when the system returns to a situation equivalent
to $t=0$ after visiting every possible path from one history to
another an equal number of times.  At $t=1$, let $00$ be the
corresponding history (without loss of generality, the random seed
history is taken to be $00$). The agents decide in a random
fashion as the history has not occurred before (or has occurred an
even number of times before). The outcome would be $1$ (or $0$)
with probability $1/2$. Let the outcome be $1$, for example. The
history bit-string will become $001$. Prior to the current $m=2$
bit-string of $01$, one history bit-string (namely $00$) occurred
once.  The strategies are now divided into $2$ ranks with

Rank-1:\qquad including strategies that predict $1$ for history
$00$

Rank-2:\qquad including strategies that predict $0$ for history
$00$

\noindent The strategy VP pattern thus consists of two ranks
corresponding to assigning $+1$ VP for those strategies in
Rank-$1$ and $-1$ for those in Rank-$2$.

If the outcome is also $1$ at $t=2$, the strategies that predict
$1$ for the history $01$ will have a higher VP. Note that the
history bit-string is now $0011$.  Prior to the current bit-string
of $11$, two $m=2$ bit-strings $00$ and $01$ occurred once.  The
strategies will then be divided into three ranks after this
timestep with

Rank-1: \qquad including strategies that predict $1$ for both
histories $00$ and $01$

Rank-2: \qquad including strategies that predict $1$ for one of
the two histories $00$ and $01$

Rank-3: \qquad including strategies that predict $0$ for both
histories $00$ and $01$

\noindent The strategy VP pattern thus consists of three ranks
corresponding to a VP of $+2$ for those strategies in Rank-$1$,
$0$ for those in Rank-$2$, and $-2$ for those in Rank-$3$.

If at some time $t$, the history $01$ happens again, i.e., the
history occurred an odd number of times prior to the one under
consideration, the outcome will be $0$ due to the crowd effect as
the outcome was $1$ in the last occurrence of the history. The
Rank-$1$ strategies will lose and the Rank-$3$ strategies will
win. As a result, the ranking of the strategies are then reduced
to two ranks with

Rank-1:\qquad including strategies that predict $1$ for history
$00$

Rank-2:\qquad including strategies that predict $0$ for history
$00$

It is important to note that for a given $m$ in the efficient
phase, there are only a {\em finite number of patterns} for the
ranking of the strategies performance. In general, we have the
following result for the strategy performance ranking pattern.

\noindent \emph{If a number of} $\kappa $\emph{\ histories
occurred an odd number of times, the strategies will be divided
into }$\kappa +1$\emph{\ ranks}.  The ranking is as follows:

Rank-1:\qquad including strategies that predicted the correct
outcome for all $\kappa $ histories concerned

Rank-2:\qquad including strategies that predicted the correct
outcome for $\kappa -1$ histories concerned

...

Rank-$\ell$:\qquad including strategies that predicted the correct
outcomes for $\kappa -\ell+1$ histories concerned

..

Rank-$\kappa +1$:\qquad including strategies that predicted the
correct outcome for $0$ histories concerned

For a given value of $m$, $0\leq \kappa \leq 2^{m}$ as there are
$2^{m}$ possible histories. In the efficient phase, while the
numbers of occurrence for every history are the same when averaged
over a long time, $\kappa$ ($\kappa=0,1,2,\dots,2^{m}$) histories
may occur an odd number of times in each timestep as the game
evolves. Therefore, the current strategy ranking pattern can be
characterized by the parameter $\kappa$.  For a timestep
corresponding to $\kappa =0$, i.e., all the histories had occurred
an even number of times, there is only one rank and all the
strategies lie in the same rank since they have tied VPs (zero
VPs).  In other words, there is only one (i.e., $C_{0}^{2^{m}}
=1$) way to achieve a ranking pattern that consists only of
Rank-$1$.

Next we deduce the probability $P(\kappa)$ of having $\kappa$
histories occur an odd number of times, without invoking too much
known details of the dynamics. Assuming that each history has
probability $1/2$ to appear as one that has occurred an odd number
of times, then out of a total of $2^{m}$ history bit-strings, the
probability
$P(\kappa )$ of having $\kappa$ histories occur an odd number of
times is
\begin{equation}
P(\kappa ) = C_{\kappa}^{2^{m}}\left( \frac{1}{2} \right)^{2^{m}}
=C_{\kappa}^{{2^{m}}}/2^{2^{m}}. \label{eq1}
\end{equation}

As the game evolves, the system maps out a path in the history
space \cite{Jefferies1}. As the game goes from one history to
another, it also makes transitions from one strategy performance
ranking pattern to another. There may be frequent ties in the
strategies' performance. A merit of the present approach is that
we take explicit account of possible ties in performance among the
strategies by grouping them into the same rank.  This is important
in the efficient phase where there are frequent tied VPs among the
strategies.

\subsection{Fraction of strategies in each rank}

The fraction of strategies in a particular rank for given value of
$\kappa$ can be calculated readily.  It turns out that the ratio
of the number of strategies in increasing ranks (recall Rank-$1$
corresponds to highest VP) follows the numbers in the Pascal
triangle. When all the histories occurred an even number of times,
there is only one rank with a fraction unity of strategies, i.e.,
all strategies, belonging to the rank.  If only one history
($\kappa =1$) occurred an odd number of times, there are two (=
$\kappa +1$) ranks with half (fraction $1/2$) of the strategies in
Rank-$1$ and the other half (fraction $1/2$) in Rank-$2$.  The
ratio of the fractions of strategies in the two ranks is $1:1$. If
two histories ($\kappa = 2$) occurred an odd number of times,
there are three (=$\kappa +1$) ranks, with a fraction $1/4$ of the
strategies in Rank-$1$, $1/2$ in Rank-$2$, $1/4$ in Rank-$3$.  The
ratio of the fractions is $1:2:1$.  For three histories occurring
an odd number of times, there are four ranks with the ratio of
fractions of strategies in the ranks given by $1:3:3:1$, and so
on.  For $\kappa$ histories occurring an odd number of times, the
fraction of strategies in Rank-$1$ is $C_{0}^{\kappa}/2^{\kappa}$,
the fraction of strategies in Rank-$2$ is
$C_{1}^{\kappa}/2^{\kappa}$, and so on. In general, the fraction
of strategies in Rank-$l$ is $C_{l-1}^{\kappa}/2^{\kappa}$, where
the denominator comes from $\sum_{i=1}^{\kappa+1} C_{i-1}^{\kappa}
= 2^{\kappa}$. The ratio of the fractions of strategies in
different ranks is thus given by $C_{0}^{\kappa}: C_{1}^{\kappa} :
\cdots : C_{l-1}^{\kappa} : \cdots : C_{\kappa}^{\kappa}$, which
are the numbers in the Pascal triangle.

\section{Number of agents using a best strategy belonging to Rank-$l$}

\subsection{Non-connected population}

Consider the case of a non-connected population, i.e. basic MG or
$p=0$ in a networked MG. As an agent uses the best-scoring strategy up
to the moment
of making decision, he will use the strategy with the lowest rank
among the $s$ strategies that he was randomly assigned at the beginning
of the game.

Let $\kappa$ be the number of histories that occurred an odd
number of times. It is convenient for later discussions to
introduce the probability
\begin{equation}
\alpha _{j}^{\kappa} \equiv \frac{1}{2^{\kappa
}}\sum_{l=j+1}^{\kappa+1}C_{l-1}^{\kappa } = \frac{1}{2^{\kappa}}
\sum_{l=j}^{\kappa} C_{l}^{\kappa}, \label{eq3}
\end{equation}
that an agent holds a strategy with performance {\em worse than}
Rank-$j$. For an agent using a Rank-$1$ strategy to decide, he
must possess at least one rank-$1$ strategy.  This happens with a
probability $1 - (\alpha_{1}^{^{\kappa}})^{s}$, where
$(\alpha_{1}^{\kappa})^{s}$ is the probability that the agent
holds $s$ strategies that are all worse than Rank-$1$.

Let $N_{l}$ be the number of agents who hold a strategy in
Rank-$l$ as their best strategy, for a given $\kappa$. In the
basic MG, this is also the number of agents who will use a
strategy in Rank-$l$ to decide their action. For a population of
$N$ agents, it follows that for given $\kappa$
\begin{equation}
N_{1}=N\left[ 1-(\alpha _{1}^{\kappa })^{s}\right]. \label{eq4}
\end{equation}
Similarly, for an agent using a Rank-$2$ strategy, he must hold
at least one Rank-$2$ strategy {\em and} must not hold any
Rank-$1$ strategy. Therefore,
\begin{equation}
N_{2}=N\left[ (\alpha _{1}^{\kappa })^{s}-(\alpha _{2}^{\kappa
})^{s}\right].  \label{eq5}
\end{equation}
In general the number of agents holding a Rank-$l$ strategy ($l =
1,2,\cdots,\kappa$) as their best strategy, is given by
\begin{equation}
N_{l}=N\left[ (\alpha _{l-1}^{\kappa })^{s}-(\alpha _{l}^{\kappa
})^{s} \right], \label{eq6}
\end{equation}
with $\alpha_{0}^{\kappa} = 1$ as given by Eq. (2). For $l = \kappa
+1$,
\begin{equation}
N_{\kappa +1}=N(\alpha _{\kappa }^{\kappa })^{s}. \label{eq7}
\end{equation}

As an example, take $s=2,$ $N=101$, and consider a moment in the
game corresponding to $\kappa =4$. Hence we have $\kappa+1 = 5$
ranks.  The ratio of strategies in these ranks is $1:4:6:4:1$. The
average number of agents using strategies in each rank in these
turns is given by $N_{1}=12.23$, $N_{2}=41.03$, $N_{3}=37.88$,
$N_{4}=9.47$, and $N_{5}=0.39$. These numbers change with time as
the game evolves to timesteps with different values of $\kappa$.
Knowing the number of agents using each rank of strategies, it is
then possible to evaluate analytically the average number of
agents making a particular decision and the mean success rate of
the agents, as we shall discuss in later sections.

\subsection{Networked Population}

Let $p$ be the probability that two randomly chosen agents are
connected. For $p\neq 0$, the agents may decide based on a
strategy that they do not hold. As a result, the number of agents
who actually {\em use} a strategy for decision in a particular
rank is, in general, {\em not} equal to the number of agents
$N_{j}$ who {\em hold} a best-scoring strategy belonging to that
rank \cite{gourley}. The number of agents $\tilde{N}_{l}(p)$ who
decide by using a Rank-$l$ strategy can formally be expressed as a
sum of two terms
\begin{equation}
\tilde{N}_{l}(p)=\overline{N}_{l}+\sum_{j=l+1}^{\kappa +1}\Delta
N_{jl}, \label{eq11}
\end{equation}
where $\overline{N}_{l}$ is the number of agents who hold a
Rank-$l$ strategy as their best-performing strategy {\em and}
are not linked to agents with a better (hence lower ranking)
performing strategy, and the second term represents all those using
a Rank-$l$ strategy due to the presence of links. Writing $q=1-p$,
$\overline{N}_{l}$ is then given by
\begin{equation}
\overline{N}_{l}=N_{l}q^{\sum_{i=1}^{l-1}N_{i}},  \label{eq9}
\end{equation}
with $N_{l}$ given by Eqs.(\ref{eq6}) and (\ref{eq7}).  In
Eq.(\ref{eq11}), $\Delta N_{jl}$ is the number of agents who hold
a Rank-$j$ strategy as their best performing strategy, but they
use a Rank-$l$ strategy for deciding because they are linked to
agents carrying such a strategy. Note that $j>l$ because an agent
will use the best performing strategy among his own strategies and
his connected neighbors' strategies in our networked MG model. Now
consider an agent who does not hold a strategy in Rank-$l$ but
{\em uses} a Rank-$l$ strategy held by one of his neighbors.  This
happens only when (i) he is not linked to any agent who holds a
strategy better than Rank-$l$ (the probability is thus
$q^{^{\sum_{i=1}^{l-1}N_{i}}}$) {\em and} (ii) he is linked to
{\em at least} one agent who holds a Rank-$l$ strategy (the
probability is $(1-q^{N_{l}})$). Hence we have
\begin{equation}
\Delta N_{jl}= N_{j} (q^{^{\sum_{i=1}^{l-1}N_{i}}}) (1-q^{N_{l}}).
\label{eq10}
\end{equation}
Equation (\ref{eq11}) for $\tilde{N}(p)$ coupled with
$\overline{N}$ is given by Eq.(\ref{eq9}), $\Delta N_{jl}$ is
given by Eq.(\ref{eq10}), and $N_{j}$ is given by Eqs.(\ref{eq6})
and (\ref{eq7}) gives the number of agents who use a strategy in
Rank-$l$ for deciding their action in a connected population.

\section{Application: Mean success rate}

\begin{figure}
\centerline{\epsfig{file=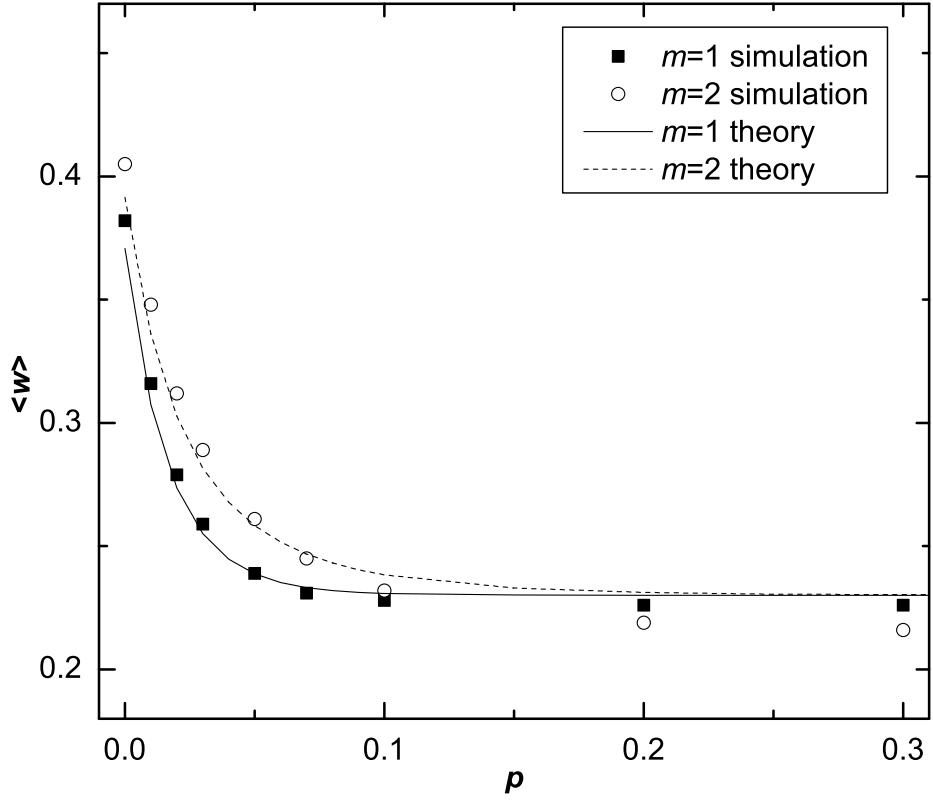,height=5in,angle=0}}
\caption{The mean success rate $\langle w \rangle$ of the agents
as a function of connectivity $p$ for $m=2$ and $m=1$. Other
parameters are $N=101$ and $s=2$.  The symbols are results
obtained by numerical simulations and the lines are analytic
results obtained by using Eq.(\ref{eq17}).} \label{figure1}
\vskip0.5in
\end{figure}

The mean success rate $\langle w \rangle$ (or mean wealth) of the
agents is the average number of winners per agent per turn. This
quantity reflects the global performance of the population as a
whole. This quantity is also closely related to the fluctuations
(or standard deviation) in the number of agents choosing a
particular option as the game proceeds.  A smaller fluctuation
implies a higher mean success rate.  Figure~\ref{figure1} shows
$\langle w \rangle$ as a function of connectivity $p$ obtained by
numerical simulations for $m=1$ and $m=2$ (symbols) in a
population of $N=101$ agents with $s=2$ strategies per agent.  As
$p$ increases, $\langle w \rangle$ decreases, together with a drop
in the spread of the success rates among the agents
\cite{gourley}. Thus in the networked MG model, while higher
connectivity ensures fairness, the efficiency also decreases.
Qualitatively, the drop in $\langle w \rangle$ comes about from
the enhanced crowd effect as $p$ increases. Here, we derive an
expression for the mean success rate as a function of connectivity
$p$. Consider a timestep $t$ corresponding to $\kappa$ histories
having occurred an odd number of times. Given this, $t$ may belong
to $\{t_{even}^{\mu}\}$ or $\{t_{odd}^{\mu}\}$ for the particular
history bit-string $\mu$ that the population is facing when making
a decision, since there are $2^{m}-\kappa$ histories which have
occurred an even number of times.


If $t\in \{t_{odd}^{\mu}\}$, the mean number of agents choosing
the last winning option of the corresponding history is given by
\begin{equation}
A_{odd}(\kappa )=\sum_{l=1}^{\kappa+1
}\tilde{N}_{l}(p)\left(\frac{\kappa -l+1}{\kappa}\right).
\label{eq12}
\end{equation}
This is because the Rank-$l$ strategies must have made the correct
predictions for $\kappa -l+1$ out of the $\kappa$ histories
concerned. Thus, the agents using a Rank-$l$ strategy have a
probability $(\kappa -l+1)/\kappa$ of choosing the previous
winning option for the history $\mu$ based on what every agent
decides.  Due to crowd effect, this is also the probability that
the agents using a Rank-$l$ strategy lose.  Therefore, they will
win with a probability $1- (\kappa-l+1)/\kappa = (l-1)/\kappa$.
The mean success rate $w_{odd}(\kappa)$ for a given $\kappa$ and
$t \in \{t_{odd}^{\mu}\}$ is
\begin{equation}
w_{odd}(\kappa )=\sum_{l=1}^{\kappa+1
}\left(\frac{\tilde{N}_{l}(p)}{N}\right)\left(\frac{l-1}{\kappa}\right).
\label{eq14}
\end{equation}

If $t\in \{t_{even}^{\mu}\}$, the agents decide randomly and the
mean number of agents choosing a particular option is $N/2$. In
this case, the probability of having $n$ agents choose a
particular option is
\begin{equation}
P_{n}=C_{n}^{N}/2^{N},   \label{eq15}
\end{equation}
as every agent has two options. For MG, the winners are those in
the minority group.  There are $n$ winners for $n < (N-1)/2$ and
$(N-n)$ winners for $n \geq (N+1)/2$.  The mean success rate
$w_{even}$ for $t \in \{t_{even}^{\mu}\}$ is then given by
\begin{equation}
w_{even}=\sum_{n=0}^{(N-1)/2}P_{n}\frac{n}{N}+\sum_{n=(N+1)/2}^{N}P_{n}
\frac{ N-n}{N}.   \label{eq16}
\end{equation}
We note that one may also make the crude approximation that
$w_{even} = 1/2$, without taking into account of the fluctuations
in the number of agents making identical decisions.

Given a value of $\kappa$, i.e., there are $\kappa$ histories
which have occurred an odd number of times and $2^{m}-\kappa$
histories which occurred an even number of times,  the probability
of having a timestep $t\in \{t_{odd}^{\mu}\}$ is $\kappa /2^{m}$.
The probability of having a timestep $t \in \{t_{even}^{\mu}\}$ is
$(1- \kappa/2^{m})$. The mean success rate $\langle w \rangle$ is
obtained by averaging over the probabilities of having $t \in
\{t_{odd}\}$ and $t \in \{t_{even}\}$ for given $\kappa$ and then
averaging over the probability of having $\kappa$ odd-occurring
strategies. The mean success rate is then formally given by
\begin{equation}
\langle w \rangle =\sum_{\kappa =0}^{2^{m}} P(\kappa )\left(
\frac{\kappa }{2^{m}} w_{odd}(\kappa) + (1-\frac{\kappa
}{2^{m}})w_{even} \right), \label{eq17}
\end{equation}
with $P(\kappa)$ given by Eq.(1). Equation (\ref{eq17}) is a
general expression for the mean success rate.  It is valid for
both non-networked and networked populations. Figure~\ref{figure1}
compares the analytic results (lines) for $\langle w \rangle$ from
Eq.(\ref{eq17}) as a function of $p$ for different values of $m=1$
and $m=3$. The results are in very good agreement with results
obtained by numerical simulations. The present formalism also
provides a physically transparent picture for the drop in $\langle
w \rangle$ with $p$.  Since there is no strategy that outperform
others, those instantaneously better performing strategy or
strategies have a higher chance of losing in immediate timesteps.
Therefore, by forcing the agents to follow the better performing
strategy of their connected neighbors actually {\em lower} their
mean success rate.  We emphasize that the present formalism is
related to the crowd-anticrowd theory \cite{johnson3,hart1} in
that the agents using a strategy and those using the corresponding
anticorrelated partner, have different success rates given by the
term in the last parentheses in Eq.(\ref{eq14}) since the pair of
strategies must belong to different rankings.

\section{Mean success rate of agents with degree $k$}

A useful way to describe the topological properties of a network
is the degree distribution, which is the distribution of the
number of connected neighbors among the nodes in a network
\cite{mendes,barabasi1}. Statistical analysis has revealed that
real world networks exhibit degree distributions of various kinds
\cite{mendes,barabasi1,Watts1,networks}. For classical random
graphs discussed in previous sections, the degree distribution is
a Poisson distribution \cite{randomgraph}; while for growing
networks with preferential attachment in its growth mechanism, the
degree distribution exhibits power law behavior
\cite{mendes,barabasi1}. While the analysis in the last section
suffice for evaluating $\langle w \rangle$ in a random network, it
will be useful to develop our formalism by focusing on agents with
a given number of neighbors, i.e. a given degree.  Here, we aim at
studying the mean success rate of agents with degree $k$ in a
networked MG.

Consider a particular agent having $k$ links to other agents.
Recall that (see Eq.(\ref{eq6})) the probability that an agent
holding a Rank-$l$ strategy as her best performing strategy is
given by $(\alpha _{l-1}^{\kappa })^{s}-(\alpha _{l}^{\kappa
})^{s}$. Note that this is also the probability that her neighbor
holds a Rank-$l$ strategy as his best performing strategy.
Combining these probabilities for an agent and his $k$ neighbors,
the probability $\gamma(\kappa,k,l)$ of an agent with $k$
neighbors using a Rank-$l$ strategy is
\begin{equation}
\gamma (\kappa,k,l)=(\alpha _{l-1}^{\kappa })^{(k+1)s}-(\alpha
_{l}^{\kappa })^{(k+1)s}.  \label{eq20}
\end{equation}
This follows from the fact that an agent who has $k$ links is
equivalent to an agent who effectively has $(k+1)s$ strategies in
hand, with repetition allowed.  Recall that the success rate or
winning probability of a Rank-$l$ strategy is $(l-1)/\kappa$ for
$t \in \{t_{odd}^{\mu}\}$. The success rate of an agent with $k$
links for timesteps $t \in \{t_{odd}^{\mu}\}$ is given by
\begin{equation}
w_{odd}(k,\kappa)=\sum_{l=1}^{\kappa +1}\gamma
(\kappa,k,l)\frac{l-1}{\kappa}.  \label{eq21}
\end{equation}
We should also take into account cases corresponding to $t \in \{
t_{even}^{\mu}\}$ for which the mean success rate of an agent is
given by $w_{even}$ in Eq.(\ref{eq16}). As a result, the success
rate of an agent with degree $k$ is given by
\begin{equation}
\langle w(k) \rangle =\sum_{\kappa =0}^{2^{m}} P(\kappa
)\left(\frac{\kappa }{2^{m}} w_{odd}(k,\kappa) + (1-\frac{\kappa
}{2^{m}})w_{even} \right). \label{eq22}
\end{equation}

For the particular case of classical random graphs, the
probability of having $k$ links in a system with $N$ nodes
(agents) for a given value of connectivity $p$ is given by
\begin{equation}
Y(k)=C_{k}^{N-1}p^{k}(1-p)^{N-1-k}.  \label{eq23}
\end{equation}
Combining with Eq.(\ref{eq22}), the mean success rate in the
population with connectivity $p$ is formally given by
\begin{equation}
\langle w \rangle =\sum_{k=0}^{N-1}Y(k) \langle w(k) \rangle.
\label{eq24}
\end{equation}

\begin{figure}
\centerline{\epsfig{file=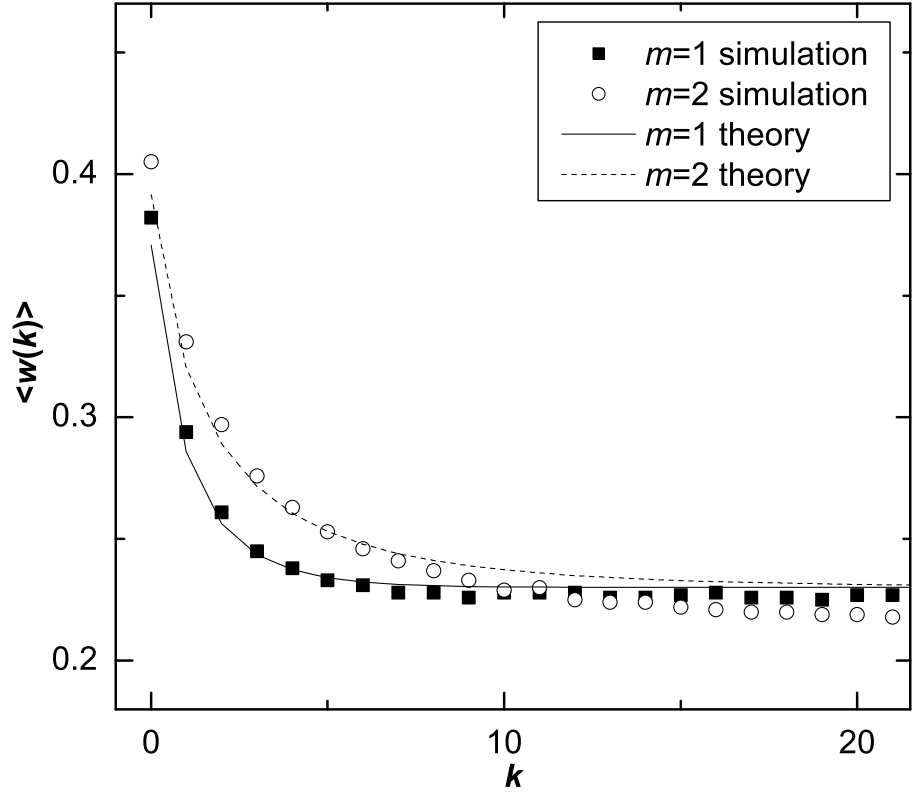,height=5in,angle=0}}
\caption{The mean success rate $\langle w(k) \rangle$ of agents of
degree $k$ as a function of $k$ for $m=2$ and $m=1$. Other
parameters are $N=101$ and $s=2$. The symbols represent numerical
results obtained by carrying simulations with the range $0 \leq p
\leq 0.5$.  The lines give the analytic results obtained by
Eq.(\ref{eq22}).} \label{figure2} \vskip0.5in
\end{figure}

Figure~\ref{figure2} shows the numerical and analytic results of
$\langle w (k) \rangle$ as a function of $k$ for $m=1$ and $m=2$.
The analytic results are, again, in good agreement numerical
results. The numerical results are obtained from data in many runs
with different values of $p$ ranging from $0 \leq p \leq 0.5$. For
a given $p$, data are obtained for values of $k$ around the mean
degree $\langle k (p) \rangle$.  We note that, for given degree
$k$ and fixed $m$, $\langle w (k) \rangle$ does not depend on $p$,
i.e., the success rate of isolated agents in a population with
$p=0.01$ is the same as that for $p=0.02$ (if isolated agents
exists).  For the present version of networked MG, the isolated
agents, i.e., those without any links, have the highest mean
success rate. This drops in $\langle w(k) \rangle$ comes about
from the fact agents with connected neighbors effectively hold a
substantial portion of the strategies and hence they will join the
crowd.  By being isolated, one can avoid the crowd and hence
achieve a higher success rate. We also checked that $\langle w
\rangle$ obtained by Eq.(\ref{eq24}) are nearly identical to those
obtained by Eq(\ref{eq23}), for small values of $m$.


\section{Discussion and extension to Networked B-A-R model}

We formulated a theory applicable to agent-based models in which a
population is competing to be in the minority group. The
population may be networked or non-networked.  The theory is based
on the tendency that the system restores itself and avoids the
existence of strategies that outperform others. This is the case
for the efficient phase in the MG.  By invoking the idea that the
strategy performance ranking patterns changes as the game evolves
and that only a finite number of patterns exist, it is possible to
study the ranking patterns based on the number of history
bit-strings that occurred an odd number of times.  The fraction of
strategies in each rank can be found, together with the number of
agents using a strategy of Rank-$l$ in order to decide. For the
case of networked populations, care must be taken to evaluate the
number of agents using a strategy of Rank-$l$ through the
connections. An expression for the mean success rate as a function
of connectivity $p$ and $m$ can be derived.  Results are found to
be in good agreement with those obtained by extensive numerical
simulations on networked MG.  A geometrical property of networks
is the degree distribution.  We derived an expression for the mean
success rate of agents for a given degree $k$ in a networked MG
with the underlying network being a classical random graph.  The
results are found to be, again, in good agreement with numerical
results. The present theory has the merit of taking into account
possible ties in the strategies' performance.

The validity of the formalism depends on the assumption that the
system passes through quasi-Eulerian paths in the history space in
the efficient phases of both non-networked and networked MG. The
details of the dynamics are not important, only that we assume the
equal probabilities of the occurrence of the possible outcomes.
The formalism can also be applied or extended to other situations
that exhibit similar features. To illustrate the idea, we consider
the interesting situation in a Binary Agent Resource (B-A-R) game
with {\em high resource level} in a {\em highly connected}
population, i.e., for high values of $p$. The B-A-R model in a
networked population represents a networked binary version of
Arthur's El Farol Problem concerning bar-attendance
\cite{arthur1,ourbook,johnson1,johnson2}.  In the B-A-R model, the
winning option is no longer decided by the minority side. Instead,
there is a general global resource level $L$ ($L < N$) which is
not announced to the agents.  At each timestep $t$, each agent
decides upon two possible options: whether to access resource $L$
(action `+1') or not. The two global outcomes at each timestep,
`resource over-used' and `resource not over-used', are denoted as
`0' and `1'. If the number of agents $n_{+1}[t]$ choosing action
$+1$ exceeds $L$ (i.e. resource over-used and hence global outcome
`0') then the $N - n_{+1}[t]$ abstaining agents win. By contrast
if $n_{+1}[t]\leq L$ (i.e. resource not over-used and hence global
outcome `1') then these $n_{+1}[t]$ agents win.

\begin{figure}
\centerline{\epsfig{file=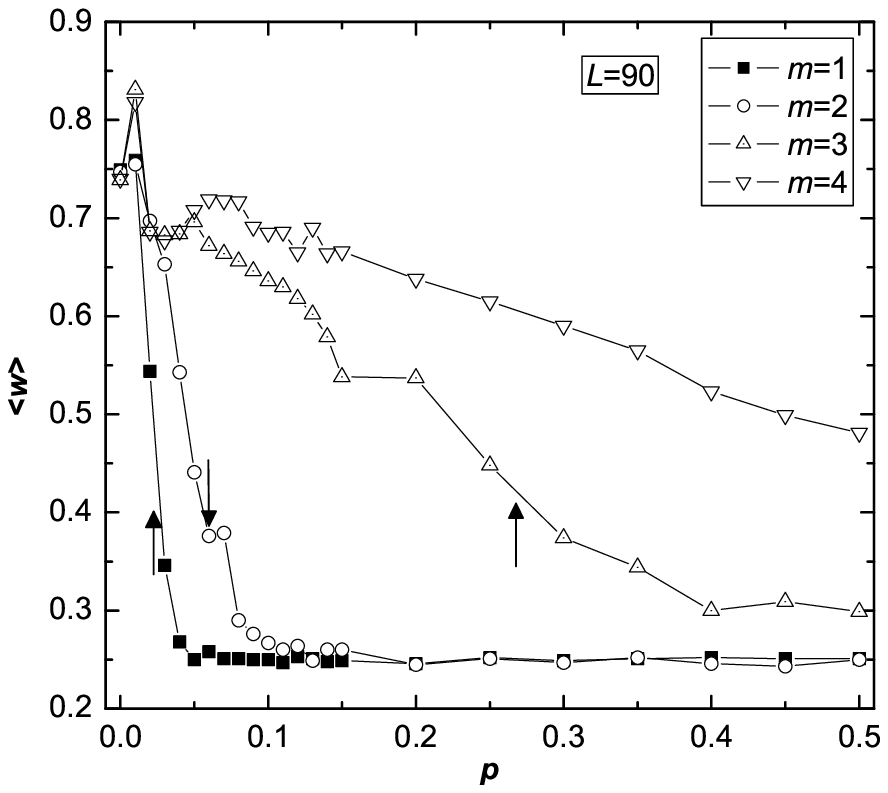,height=5in,angle=0}}
\caption{The mean success rate $\langle w \rangle$ as a function
of connectivity $p$ in the B-A-R model at high resource level
($L=90$) for $m=1,2,3,4$ obtained by numerical simulations. The
lines are guides to eye. Other parameters are $N=101$ and $s=2$.
The arrows indicate the estimate of $p_{c}(m)$ using
Eq.(\ref{eq19}) above which the system goes into a high-$p$
state.} \label{figure3} \vskip0.5in
\end{figure}

Numerical results for a high resource level B-A-R model show
interesting features as a function of the connectivity $p$.
Figure~\ref{figure3} shows the dependence on the mean success rate
for $L=90$ in a $N=101$ population as a function of $p$. For
$L>3N/4$ in a non-connected population ($p=0$), the system is in a
frozen state in the sense that the outcome is persistently $1$,
i.e., the resource is persistently not over-used with $3N/4$
winners per turn.  It is observed that as $p$ increases, the
system moves away from the frozen state \cite{gourley}.  A
high-$p$ limit is eventually reached corresponding to a state of
anti-persistence or double periodicity, characterized by an
outcome time series with equal probability for the two possible
outcomes and a mean success rate slightly lower than $1/4$. In
particular, it is observed that the value of $p$ (denoted by
$p_{c}(m)$) above which the system reaches the high-$p$ state,
depends sensitively on $m$ and increases with $m$.


The present theory can be extended to estimate $p_{c}(m)$. To
proceed, we propose a criteria that the system is anti-persistent
only if $A_{odd}(\kappa)
> L$ for {\em all} $\kappa$. This can be understood easily since
anti-persistence implies that for $t \in \{t_{odd}^{\mu}\}$ for
the history $\mu$ concerned, the outcome will be opposite to that
in the last occurrence of the history. However, in B-A-R model,
for $A_{odd}(\kappa)<L$, the winning option in the last occurrence
of the history wins again, and the system ceases to be
anti-persistence. We further note that $A_{odd}(\kappa)$ is a
monotonically decreasing function of $\kappa$, with a minimum at
$\kappa = 2^{m}$ when all the possible histories occurred an odd
number of times.  This behavior follows from Eq.(\ref{eq12}), and
we have also checked it against numerical results.

For a given high resource level $L$, as $p$ decreases from the
high-$p$ state, 
the difference between $A_{odd}(\kappa)$ and $L$ drops. Eventually
when $A_{odd}(\kappa )<L$, the system is no longer anti-persistent
for some $\kappa$. As $A_{odd}(\kappa)$ takes on its minimum value
at $\kappa=2^{m}$, an estimate on the breakdown of anti-persistent
behavior is then given by the condition
\begin{equation}
A_{odd}(2^{m})=L.  \label{eq19}
\end{equation}
Equation (\ref{eq19}) can be used to estimate the critical value
$p_{c}(m)$ for fixed resource level. To test the validity, we take
a system of $N=101$, $L=90$, and $s=2$ (see Fig.~\ref{figure3}).
The values of $p_{c}$ turns out to be $p_{c} = 0.0220$ for $m=1$,
$p_{c} = 0.0592$ for $m=2$, $p_{c}= 0.2738$ for $m=3$, and
$p_{c}=0.9988$ for $m=4$, as marked by the arrows in
Fig.\ref{figure3}. The results capture the trend that $p_{c}(m)$
increases with $m$. For $m=5$, our estimate shows that the system
cannot achieve an anti-persistent high-$p$ state even if $p=1$, a
result again consistent with numerical results \cite{gourley}.
Similarly, one may vary the resource level $L$ at given $p$ and
Eq.(\ref{eq19}) can be used to estimate the critical resource
level $L_{c}(m)$ above which the system starts to deviate from an
anti-persistent state.

The formalism can also be applied to a non-networked B-A-R game
with resource level $L \gtrsim N/2$, for which the outcome series
and thus the history series also exhibit anti-persistence or
doubly periodic features \cite{zheng2}. The formalism can also be
extended to cases in which the outcome (hence history) time-series
shows known features other than anti-persistence. For example, it
can be modified to study each of the many states that a high
resource level B-A-R game passes through from the frozen state at
$p=0$ to the high-$p$ state as the connectivity varies. The
starting point is to give a known outcome series, e.g.,
$11101110\dots$ for $m=1$ and $11111101111110\dots$ for $m=2$ just
off the frozen state. Once the pattern of history time series is
known, the part of the full history space that matters is also
known and thus the ranking pattern of the strategies can be worked
out \cite{sonic1}. Similar situations also happen in the Networked
B-A-R model \cite{gourley} with high resources \cite{gourley2}.
The formalism can also be extended to study different variations
on the basic MG, such as the thermal MG \cite{thermal1, thermal2}
and MG with biased strategy pools \cite{larry2}; and to different
versions of networked MG in which neighboring agents compare their
wealth instead of strategy performance \cite{LosAlamos,Hui}.

\begin{acknowledgements}
This work was supported in part by the Research Grants Council of
the Hong Kong SAR Government under grant number CUHK 4241/01P. We
would like to thank S. Gourley, S.C. Choe, and K.P. Chan for
useful discussions.

\end{acknowledgements}

\newpage \centerline{\bf Figure Captions}

\bigskip

\noindent Figure 1: The mean success rate $\langle w \rangle$ of
the agents as a function of connectivity $p$ for $m=2$ and $m=1$.
Other parameters are $N=101$ and $s=2$.  The symbols are results
obtained by numerical simulations and the lines are analytic
results obtained by using Eq.(\ref{eq17}).

\bigskip

\noindent Figure 2: The mean success rate $\langle w(k) \rangle$
of agents of degree $k$ as a function of $k$ for $m=2$ and $m=1$.
Other parameters are $N=101$ and $s=2$. The symbols represent
numerical results obtained by carrying simulations with the range
$0 \leq p \leq 0.5$.  The lines give the analytic results obtained
by Eq.(\ref{eq22}).

\bigskip

\noindent Figure 3: The mean success rate $\langle w \rangle$ as a
function of connectivity $p$ in the B-A-R model at high resource
level ($L=90$) for $m=1,2,3,4$ obtained by numerical simulations.
The lines are guides to eye. Other parameters are $N=101$ and
$s=2$. The arrows indicate the estimate of $p_{c}(m)$ using
Eq.(\ref{eq19}) above which the system goes into a high-$p$ state.

\newpage

\newpage

\newpage


\newpage


\begin{thebibliography}{99}
\bibitem{recentactivities} For an overview of recent progress and
activites in agent-based modelling of complex systems, see, for
example, http://sbs-xnet.sbs.ox.ac.uk/complexity/ and
http://www.ima.umn.edu/complex/.

\bibitem{ourbook} N.F. Johnson, P. Jefferies, and P.M. Hui,
{\em Financial Market Complexity} (Oxford University Press, 2003).

\bibitem{arthur1} B. Arthur, Amer. Econ. Rev. {\bf 84}, 406 (1994);
Science {\bf 284}, 107 (1999).

\bibitem{johnson1} N.F. Johnson, S. Jarvis, R. Jonson, P. Cheung,
Y.R. Kwong, and P.M. Hui, Physica A {\bf 258} 230 (1998).

\bibitem{challet1} D. Challet and Y.C. Zhang, Physica A {\bf 246}, 407
(1997);
{\em ibid.} {\bf 256}, 514 (1998).

\bibitem{challet0} See the website
http://www.unifr.ch/econophysics/minority for updated information
on the minority game.

\bibitem{johnson2} For a review, see N.F. Johnson and P.M. Hui,
cond-mat/0306516.

\bibitem{zheng2} N.F. Johnson, P.M. Hui, D. Zheng, and C.W. Tai,
Physica A {\bf 269}, 493 (1999).

\bibitem{challet7} D. Challet, M. Marsilli, and G. Ottino,
cond-mat/0306445.

\bibitem{savit} R. Savit, R. Manuca, and R. Riolo, Phys. Rev. Lett.
{\bf 82}, 2203 (1999).

\bibitem{challet3} D. Challet, M. Marsili, and R. Zecchina, Phys. Rev.
Lett. {\bf 85}, 5008 (2000).

\bibitem{challet4} D. Challet, M. Marsili, and Y.C. Zhang,
Physica A {\bf 294}, 514 (2001).

\bibitem{johnson3} N.F. Johnson, M. Hart, and P.M. Hui, Physica A
{\bf 269}, 1 (1999).

\bibitem{hart1} M. Hart, P. Jefferies, N.F. Johnson, and P.M. Hui,
Physica A {\bf 298}, 537 (2001).

\bibitem{challet5} D. Challet and M. Marsili, Phys. Rev. E {\bf 60},
R6271 (1999).

\bibitem{challet6} M. Marsili, D. Challet, and R. Zecchina,
Physica A {\bf 280}, 522 (2000).

\bibitem{coolen1} J.A.F. Heimel and A.C.C. Coolen, Phys. Rev. E
{\bf 63}, 056121 (2001).

\bibitem{coolen2} J.A.F. Heimel, A.C.C. Coolen, and D.
Sherrington, Phys. Rev. E {\bf 65}, 016126 (2001).

\bibitem{challet8} D. Challet and M. Marsili, Phys. Rev. E {\bf
62}, 1862 (2000).

\bibitem{galla} T. Galla, preprint cond-mat/0406158.

\bibitem{gourley} S. Gourley, S.C. Choe, N.F. Johnson, and P.M.
Hui, preprint cond-mat/0401527.

\bibitem{choe} S.C. Choe, N.F. Johnson, and P.M. Hui, preprint
cond-mat/0405037.

\bibitem{mendes} S.N. Dorogovtsev and J.F.F. Mendes, {\em Evolution
of Networks: From Biological Nets to the Internet and WWW} (Oxford
University Press, Oxford, 2002); Adv. Phys. {\bf 51}, 1079 (2002).

\bibitem{larry1} K. F. Yip, T.S. Lo, P.M. Hui, and N.F. Johnson,
Phys. Rev. E {\bf 69}, 046120 (2004).

\bibitem{zheng1} D. Zheng and B.H. Wang, Physica A {\bf 301}, 560
(2001).

\bibitem{Jefferies1} P. Jefferies, M.L. Hart, and N.F. Johnson,
Phys. Rev. E {\bf 65}, 016105 (2002).

\bibitem{barabasi1} R. Albert and A.-L. Barab\'{a}si, Rev. Mod.
Phys. {\bf 74}, 47 (2002).

\bibitem{Watts1} D.J. Watts, {\em Six Degrees: The Science of a
Connected Age}
(William Heinemann, London, 2003).

\bibitem{networks} D.J. Watts, S.H. Strogatz, Nature {\bf 393}, 440
(1998);
B.A. Huberman, L.A. Adamic, Nature {\bf 401}, 131 (1999); A.-L.
Barab\'{a}si, R. Albert, H. Jeong, G. Bianconi, Science {\bf 287},
2115 (2000); F. Liljeros, C.R. Edling, L.A. N. Amaral, H.E.
Stanley, Y. \AA berg, Nature {\bf 411}, 907 (2001); M.E.J. Newman,
S.H. Strogatz, D.J. Watts, Phys. Rev. E {\bf 64}, 026118 (2001);
S. Maslov and K. Sneepen, Science {\bf 296}, 910 (2002).

\bibitem{randomgraph} B. Bollob\'{a}s, {\em Random Graphs}
(Academic Press, London, 1985).

\bibitem{sonic1} H.Y. Chan, T.S. Lo, P.M. Hui, and N.F. Johnson
(unpublished).

\bibitem{gourley2} S. Gourley, H.Y. Chan, P.M. Hui, and N.F.
Johnson (unpublished).

\bibitem{thermal1} A. Cavagna, J.P. Garraham, I. Giardina, and
D. Sherrington, Phys. Rev. Lett. {\bf 83}, 4429 (1999).

\bibitem{thermal2} M.L. Hart, P. Jefferies, N.F. Johnson, and P.M. Hui,
Phys. Rev. E {\bf 63}, 017102 (2001).

\bibitem{larry2} K.F. Yip, P.M. Hui, T.S. Lo, and N.F. Johnson,
Physica A {\bf 321}, 318 (2003).

\bibitem{LosAlamos} M. Anghel, Z. Toroczkai, K.E. Bassler, G.
Kroniss, Phys. Rev. Lett. {\bf 92}, 058701 (2004).

\bibitem{Hui} T.S. Lo, K.P. Chan, P.M. Hui, and N.F. Johnson
(unpublished).

\end{thebibliography}
\end{document}